\documentclass[prl, aps, longbibliography, twocolumn, amsmath,amssymb]{revtex4-1} 
\usepackage{graphicx} 
\usepackage{hyperref} 
\usepackage{color}
 
\begin{document}

\title{Cavity mediated manipulation of  distant spin currents using cavity-magnon-polariton}

\author{Lihui Bai}
\email{bai@physics.umanitoba.ca}
\affiliation{Department of Physics and Astronomy, University
of Manitoba, Winnipeg, Canada R3T 2N2}

\author{Michael Harder }

\affiliation{Department of Physics and Astronomy, University
of Manitoba, Winnipeg, Canada R3T 2N2}

\author{Paul Hyde }

\affiliation{Department of Physics and Astronomy, University
of Manitoba, Winnipeg, Canada R3T 2N2}

\author{Zhaohui Zhang}

\affiliation{Department of Physics and Astronomy, University
of Manitoba, Winnipeg, Canada R3T 2N2} 

\author{Y. P. Chen}
\affiliation{Department of Physics and Astronomy, University of Delaware, Newark, Delaware 19716, USA}

\author{John Q. Xiao}
\affiliation{Department of Physics and Astronomy, University of Delaware, Newark, Delaware 19716, USA}

\author{Can-Ming Hu}

\affiliation{Department of Physics and Astronomy, University
of Manitoba, Winnipeg, Canada R3T 2N2}

\date{\today}

\begin{abstract} 
Using electrical detection of a strongly coupled spin-photon system comprised of a microwave cavity mode and two magnetic samples, we demonstrate the long distance manipulation of spin currents.  This distant control is not limited by the spin diffusion length, instead depending on the interplay between the local and global properties of the coupled system, enabling systematic spin current control over large distance scales (several centimeters in this work).  This flexibility opens the door to improved spin current generation and manipulation for cavity spintronic devices. 

\end{abstract}

\maketitle

In spintronic devices information is carried by spin current, rather than charge current, and therefore information processing requires precise spin current manipulation.
For this purpose Datta and Das \cite{Datta1990} proposed the spin field-effect-transistor in which spin current is manipulated by a gate voltage via a local spin-orbit interaction in a semiconductor channel \cite{Nitta1997}.
On the other hand the exchange interaction is also commonly used to manipulate spin current.
For example, the production of spin current can be realized via the spin-polarization of a charge current in ferromagnetic materials and spin current can also be absorbed by a local magnetization through spin transfer torque \cite{R.H.Silsbee1979, Slonczewski1996, Berger1996, Stiles2002}.
Devices exploiting either spin-orbit or exchange interactions for spin current control are typically limited by the $\sim \mu$m spin diffusion length which depends on the spin-flip scattering time.
Although this is much larger than the $\sim$ nm range of the interactions themselves, a long distance ($\gg \mu$m) spin manipulation would be beneficial for spintronic applications.
For example, the microwave power generated in spin valves due to spin transfer torque driven dynamics \cite{Kaka2005} could be greatly enhanced using a long distance spin-interaction by enabling phase-locking of several spin systems.

In the field of cavity quantum electrodynamics the correlation of two distant "atomic resonators" has already been demonstrated using long range photon mediated interactions \cite{DeVoe1996,Mlynek2014, VanLoo2013, Reimann2015}.
Such interactions are typically on the order of the photon wavelength and are therefore much larger than those of either spin-orbit or exchange interactions, approaching the limit of the spin diffusion length in optical systems with much larger correlations possible in microwave analogues.
The recent observation of strong magnon-photon coupling in ferromagnetic/microwave cavity structures \cite{Huebl2013,Tabuchi2014,Zhang2014,Goryachev2014,Bai2015} opens the door to apply such photon mediated interactions in magnetic systems to manipulate the magnetization and spin current \cite{Bai2016,Tabuchi2015, Zhang2015b}.
 
In this work, we experimentally studied the microwave mediated interaction between two magnetic systems, demonstrating spin current manipulation over a distance of several cm. 
Using an electrical detection technique we are able to locally detect the spin currents in each magnetic system via the spin pumping effect \cite{Bai2015}.
Although the cooperativity of only one magnetic system was controlled, we find a simultaneous change in the spin current of another magnetic system which is well separated and not directly tuned. In this sense, we realized the manipulation of distant spin currents using the cavity-magnon-polariton.
Control of the cooperativities is the key to such a cavity-mediated interaction and a coupling model including both magnetic samples and a cavity mode is used to clearly highlight the effect of each photon-magnon cooperativity and to interpret the experimental observations. 
This work offers a new way to coherently control spin current and magnetization dynamics both directly and over long distances, which we expect to play an important role in the development of cavity spintronics.

\begin{figure}[tbp]
\includegraphics[width = 8.50 cm]{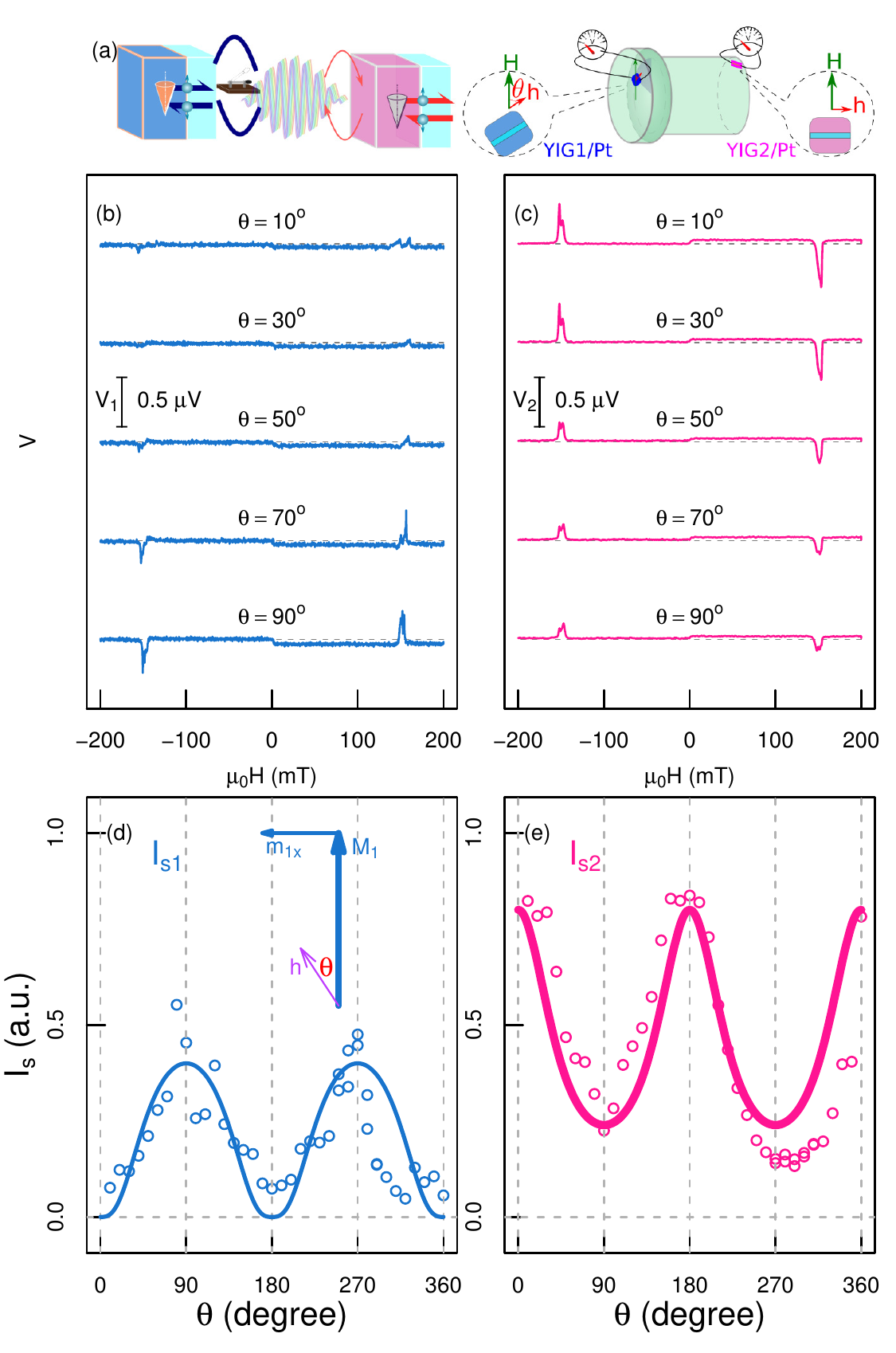} 
\caption{(Color online) Directly controlled spin current from YIG1 and long distance manipulation of spin current from YIG2. (a) Two YIGs coupled to a microwave cavity.  By tuning the YIG1 cooperativity, the YIG1 spin current was tuned. (b) The YIG1 voltage signal depends directly on the angle $\theta$, which controls cooperativity of YIG1, while (c) the voltage on YIG2 is also tuned by $\theta$. The spin currents from YIG1 and YIG2 as a function of angle $\theta$ are summarized in (d) and (e) respectively.   } 
\label{fig1}
\end{figure}

In our experiment we chose two pieces of yttrium iron garnet (YIG) on GGG substrates as the two magnetic systems due to their low Gilbert damping and low eddy current dissipation.
The two nearly identical YIG samples had dimensions of 10 mm $\times$ 7 mm $\times$ 2.6 $\mu$m, a saturation magnetization of $\mu_{0} M_{s}$ = 160 mT, a Gilbert damping of $\alpha =  3.6 \times 10^{-4}$ and a Gyromagnetic ratio of $\gamma$ =  27.6$\times2\pi \mu_{0}$GHz/T.
An externally applied magnetic field {\bf H} determined the ferromagnetic resonance (FMR) frequency according to the Kittel equation $\omega_{r}$ as $\omega_{r} = (\omega_{0}^{2} + \omega_{0}\omega_{m})^{1/2}$.
Here, $\omega_{0} =  \gamma H$ and $\omega_{m} = \gamma M_{0}$.
The microwave magnetic field ({\bf h}) driven magnetization ($M$) precession, governed by the Landau-Lifshitz-Gilbert (LLG) equation,
produced a non-equilibrium magnetization which generated a spin current through diffusion \cite{Tserkovnyak2005}. 
For the electrical detection of this spin current, platinum (Pt) strips were deposited on top of each piece of YIG with a dimension of 10 mm $\times$ 1 mm $\times$ 10 nm.  In the Pt strips the spin currents were electrically detected by conversion into charge currents via the inverse spin Hall effect (ISHE).
The spin current $I_{s}$ pumped by each sample is linearly proportional to the voltage $V_{SP}$ detected via the ISHE, $V_{SP} \propto I_{s} \propto Im(m_{x}^{*}m_{y})$ \cite{Jiao2013} where
$m_{x}$ and $m_{y}$ are the dynamical components of the magnetization in each sample. 
 
To couple the magnetic samples, a cylindrical microwave cavity made of oxygen-free copper was fabricated with a diameter of 36 mm and a height of 10 mm.
The cavity is designed to have a TM${}_{011}$ mode at $\omega_{c}/2\pi$ = 6.34 GHz.  The TM${}_{011}$ mode is chosen due to its well suited field profile with a microwave electric field along the cylindrical axis and a microwave magnetic field surrounding the electric field flux.
This mode has an unloaded damping of $\beta$ = 0.0003 (Q = 1670).
Denoting the microwave magnetic field by $h$ and the driving microwave amplitude by $h_{0}$, the cavity mode frequency profile can be written as $h = \omega^{2}h_{0}/(\omega^{2} - \omega_{c}^{2} + 2i\beta\omega_{c}\omega )$. 
Here, $\omega$ is the microwave frequency.

One of the YIG/Pt samples (labelled as YIG1) was placed on the lid of the cavity while another YIG/Pt sample (labelled as YIG2) was fixed on the bottom.
The external magnetic field {\bf H} was applied in-plane for both magnetic samples and perpendicular to the Pt strip of YIG2 in order to detect the maximum spin pumping signal via the ISHE.
The position of YIG1 with respect to the microwave magnetic flux inside of the cavity was controlled by rotating the lid with the angle $\theta$ denoting the angle between the external magnetic field {\bf H} and the local microwave magnetic field {\bf h} at YIG1 as shown in Fig. \ref{fig1} (a).  
With only YIG1 loaded, the cavity mode frequency was red shifted by 3\% to 6.155 GHz while the damping increased to $\beta = 0.0018$ (Q = 280).  With both samples loaded and wired out for electrical detection the cavity mode was red shifted by another 3\% to $\omega_{c}$ = 5.960 GHz with a damping of $\beta = 0.0052$ (Q = 100).
Meanwhile, tuning the angle $\theta$ changed the cavity mode frequency by less than 1$\%$ (much less than the shift due to loading both samples) and therefore this shift is not considered in the detection and calculation of the spin currents.
The coupling between the YIG samples and the cavity mode was characterized by measuring the microwave transmission using a vector network analyzer (VNA) \cite{Huebl2013, Tabuchi2014, Zhang2014, Goryachev2014, Bai2015} while electrical detection was performed using a lock-in technique with frequency modulation of 8.33 kHz \cite{Bai2015} with spin pumping voltages on both YIG1 and YIG2 measured simultaneously.

Spin currents were detected on both YIG1 and YIG2 by sweeping the magnetic field {\bf H} at a microwave frequency of 6 GHz, slightly detuned from the cavity mode frequency $\omega_{c}$.
The microwave output power was 100 mW. 
As shown in Fig. \ref{fig1} (b) and (c), the voltage signals have Lorentz line shapes and are anti-symmetric about the magnetic field {\bf H} as expected for spin pumping voltages \cite{Bai2013}. 
Rotating the angle $\theta$ from $0^{\circ}$ to $90^{\circ}$ by rotating YIG1, the torque exerted on the magnetization of YIG1 by the local microwave magnetic field was significantly enhanced.
Consequently, the amplitude of the spin current pumped by YIG1 is increased as shown in Fig. \ref{fig1} (b).
Simultaneously the spin current pumped by YIG2 was also detected as $\theta$ was tuned.  
Figure \ref{fig1} (c) shows a clear systematic change of the spin current pumped by YIG2, even though YIG2 was spatially separated from YIG1 and was {\it not} tuned directly. 
Contrary to the increase of spin current in YIG1, the amplitude of spin current from YIG2 decreases as $\theta$ is increased from $0^{\circ}$ to $90^{\circ}$. 

By adopting a Pt spin Hall angle of 0.0023 \cite{Bai2013} to determine the ISHE coefficient, the detected voltages were converted to spin currents.
The spin currents measured from both magnetic samples are summarized in Fig. \ref{fig1} (d) and (e) respectively.  These panels systematically demonstrate the $\theta$-dependence of the spin pumping voltages induced by the ISHE.  Note that since the external magnetic field {\bf H} was not rotated during the measurement, the spin currents from both samples maintain the same sign.   
We found that YIG1 and YIG2 spin currents are both strongly dependent on the angle $\theta$ with $I_{s1}$ having a $|\sin\theta|$ dependence with the inverse behaviour shown by $I_{s2}$; that is a maximum $I_{s1}$ signal will correspond to a minimum $I_{s2}$ signal and vice versa. 
Thus Fig. 1 illustrates the key experimental features demonstrating a long distance manipulation of the spin current on YIG2 by controlling the coupling between YIG1 and the cavity mode.

\begin{figure}[tbp]
\includegraphics[width = 8.20 cm]{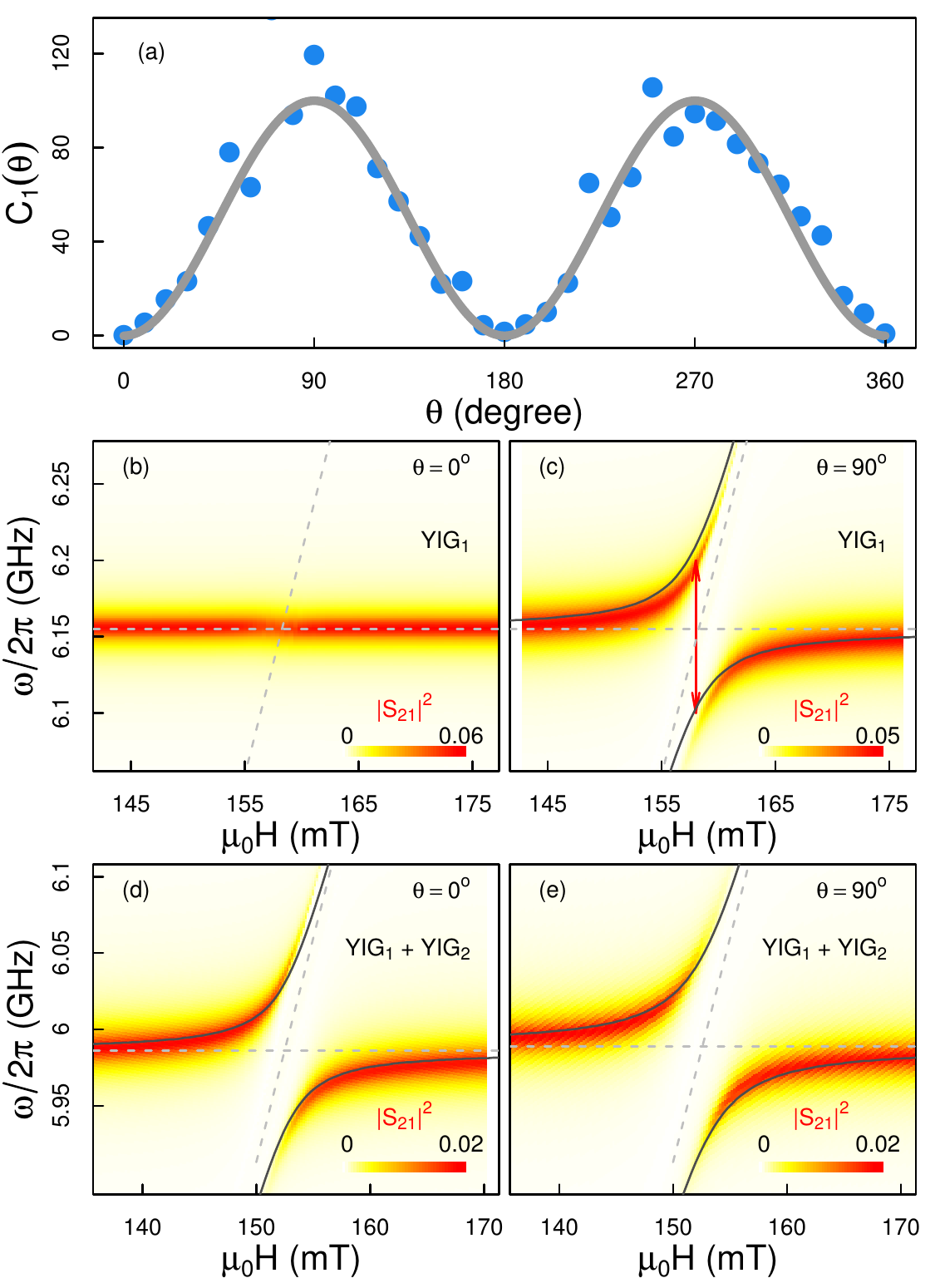} 
\caption{(Color online) (a) The cooperativity in a single YIG and cavity system is measured as a function of $\theta$ with a calculation according to Eq. \eqref{eq1_C101}. A microwave transmission $S_{21}$ measurement of the $\omega-H$ dispersion is plotted for $\theta = 0^{\circ}$ and $90^{\circ}$ in (b) and (c) respectively. (d) and (e) display the transmission for the cavity mode with both YIG1 and YIG2 loaded at YIG1 angles of $\theta = 0^{\circ}$ and $90^{\circ}$ respectively.   } 
\label{fig2}
\end{figure}

The solid curves which are shown in Fig. \ref{fig1} (d) and (e) have been calculated using a model of strongly coupled cavity-magnon-polaritons. 
To understand the manipulation of distant spin currents, we may start by understanding the controllable coupling between one YIG and a cavity mode.
In the linear coupling regime all models of such strongly coupled systems reduce to that of two coupled oscillators, one representing the cavity mode and the other YIG FMR mode, with coupling strength $\kappa_{1}$ \cite{Huebl2013, Tabuchi2014, Zhang2014, Bai2015}.
By defining the detuning parameters $\Delta_{c} \equiv (\omega^{2} - \omega_{c}^{2})/(2\beta\omega_{c}\omega)$ and $\Delta_{r} \equiv (\omega^{2} - \omega_{r}^{2})/(2\alpha\omega_{r}\omega)$ and a cooperativity $C_{1} \equiv \kappa_{1}^{2}\omega_{m}/(4\alpha\beta\omega_{c})$, the normal mode dispersion and spin current are, respectively,
\begin{subequations}
\begin{align}
\Delta_{c}\Delta_{r} &=  1 + C_{1} \label{eq1_C100} \\
I_{s1} &\propto \frac{C_{1}}{(\Delta_{c} + \Delta_{r})^{2}}\label{eq1_I1}
\end{align}
\label{eq1}
\end{subequations}
Strong coupling is defined as $C_{1} > 1$ which physically means that the rate of energy exchange between the magnetic and cavity subsystems is greater than the damping of each subsystem.
Eq. \eqref{eq1_C100} determines the $\omega-H$ dispersion and displays an anti-crossing when the cooperativity is nonzero, showing that the normal modes of the FMR/cavity system are only supported when both detunings, $\Delta_{c}$ and $\Delta_{r}$, are inversely proportional.
The key to our technique for long distance control is that the dispersion determined by Eq. \eqref{eq1_C100} is a $global$ property of the coupled system (which can be measured through both VNA and electrical techniques) while the spin current of Eq. \eqref{eq1_I1} depends both on the $local$ properties through the cooperativity $and$ the $global$ properties through the detunings $\Delta_{c}$ + $\Delta_{r}$.
Furthermore the spin current can be $locally$ measured through electrical detection, enabling multiple samples to be individually detected.   

The spin current can be directly controlled by tuning the cooperativity and since the cooperativity depends on the filling factor, this can be done by either changing the number of spins in the magnetic material \cite{Huebl2013, Tabuchi2014} or changing the local field distribution \cite{Zhang2014, Bai2015}.
By rotating the sample we can change the microwave magnetic field torque on the magnetization and therefore

\begin{equation}
C_1 = C_{1}(\theta) \propto  \sin^{2}\theta  \label{eq1_C101} 
\end{equation}
Experimental results of the cooperativity as a function of angle $\theta$ are summarized in Fig. \ref{fig2} (a) (solid circles) and compared to the prediction of Eq. \eqref{eq1_C101} (solid curve). 
Figures \ref{fig2} (b) and (c) display the global $\omega$-H dispersion in a microwave transmission spectrum $S_{21}$. 
When the cooperativity $C_{1}$ is close to 0, at $\theta = 0^{\circ}$, the maximum amplitude of $|S_{21}|^{2}$ remains at the uncoupled cavity mode frequency $\omega_{c}$ for all $H$, indicating the diminished coupling in this configuration. 
However when the cooperativity is increased by setting $\theta = 90^{\circ}$, a clear anti-crossing feature is observed near the crossing of the uncoupled cavity mode and the FMR dispersion, denoted by dashed lines respectively.
This dispersion agrees well with the solid curves which are calculated based on Eq. \eqref{eq1_C100}. 
With both magnetic samples loaded, the microwave transmission $S_{21}$ in Fig. \ref{fig2} (d) and (e) again shows that the dispersion can be tuned by changing $\theta$ for YIG1.
This illustrates how the global properties of the coupled system will depend on the local tuning of one ferromagnetic sample.
We emphasize that the $\omega-H$ dispersion can be measured using microwave transmission as shown here, or using electrical detection.
However, the electrical detection method also allows us to locally detect the spin current of individual magnetic samples which is not possible through microwave transmission. 

When two samples with a nearly identical resonance response are placed inside the cavity, we can write the normal mode dispersion and the spin current in YIG1 and YIG2 as, respectively,

\begin{subequations}
\begin{align}
\Delta_{c}\Delta_{r} &= 1 + C_{1}(\theta) + C_{2} \label{eq2_dispersion} \\
I_{s1} &\propto \frac{C_{1}(\theta)}{(\Delta_{c} + \Delta_{r})^{2}}\label{eq2_I1}\\
I_{s2} &\propto \frac{C_{2}}{(\Delta_{c} + \Delta_{r})^{2}} \label{eq2_I2}
\end{align}
\label{eq2}
\end{subequations}

Intuitively, as indicated by Eq. \eqref{eq2_dispersion}, the dispersion of two magnetic samples (with the same FMR frequency) coupled to a cavity mode differs from Eq. \eqref{eq1_C100} only by the sum of the cooperativities.  This pattern holds for any number of magnetic samples. 
This feature explains the coupling strength enhancement between Fig. \ref{fig2} (d) and (e) due to the increased number of spins.  
Furthermore, the amplitude of the spin current produced in each magnetic sample, given by Eqs. \eqref{eq2_I1} and \eqref{eq2_I2} follows the same structure as Eq. \eqref{eq1_I1}.  The only implicit difference is that the detunings now satisfy a modified dispersion depending on all cooperativities and therefore differ from the case of a single YIG sample.

\begin{figure}[tbp]
\includegraphics[width = 8.20 cm]{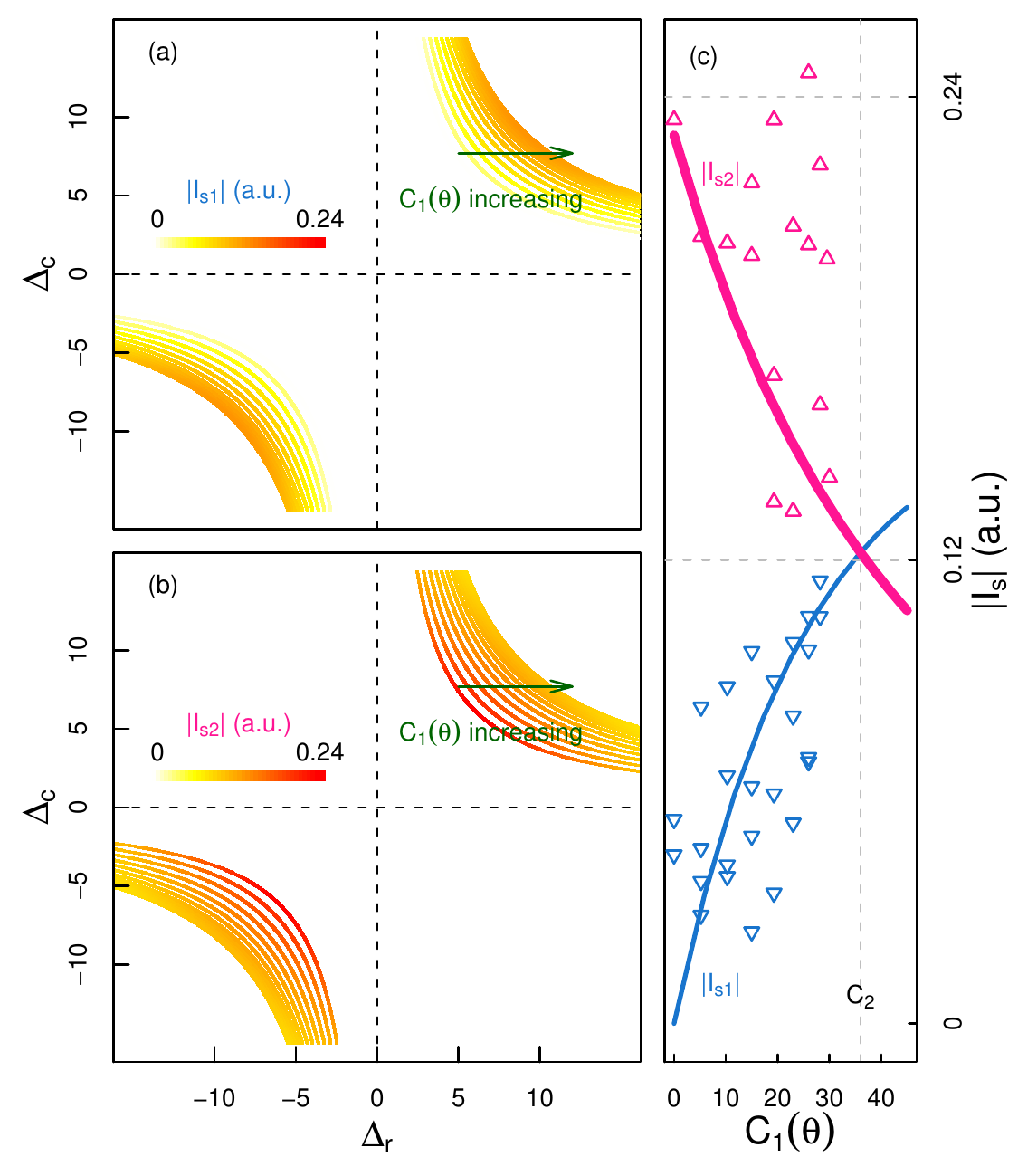} 
\caption{(Color online) Following Eq. \ref{eq2}, the dispersion and amplitudes of both spin currents are plotted in (a) and (b) as a function of cooperativity $C_{1}(\theta)$. The color scales indicate the amplitudes of both spin currents $I_{s1}$ and $I_{s2}$ while the green arrows indicate the change of $C_{1}(\theta)$ by tuning the angle $\theta$. Along the arrows at a given cavity frequency detune $\Delta_{c}$, the amplitudes of both spin currents $I_{s1}$ and $I_{s2}$ are plotted as a function of the $C_{1}$ in (c). } 
\label{fig3}
\end{figure}

Figure \ref{fig3} (a) and (b) shows the cavity-FMR detuning ($\Delta_{c}-\Delta_{r}$) dispersion following Eq. \eqref{eq2_dispersion} with the color gradient indicating the spin current amplitudes for $I_{s1}$ and $I_{s2}$ respectively.  The difference in color gradient between panels (a) and (b) highlights the difference between the direct and long distance tuning respectively.
Two normal modes are only excited when the detunings are either both positive or both negative.  Based on Fig. \ref{fig3} (a) and (b) we can summarize the spin current features in such a coupled system as follows:
(1) The normal modes of the coupled system rely on the sum of cooperativities of all magnetic samples with the coupling strength increasing when more magnetic samples are added.  
(2) The spin current pumped by each magnetic sample depends on both the global properties of the normal mode detunings and the local cooperativity with the cavity mode. 
(3) The amplitude of the spin current from YIG1 (the directly tuned sample) is increased by increasing $C_{1}$ while the amplitude of spin current from YIG2 (the distant sample) is decreased by increasing $C_{1}$.
For comparison with the experimental observation where we measured the spin current by sweeping the magnetic field at a given microwave frequency (a fixed cavity mode frequency detuning of $\Delta_{c} = 7.7$), an arrow in both Fig.\ref{fig3} (a) and (b) indicates the direction of cooperativity tuning as $\theta$ is increased from $0^{\circ}$ to $90^{\circ}$ .
Along the arrow the amplitudes of both spin currents are plotted as a function of the $C_{1}$ in (c).
The solid curves in Fig.\ref{fig1} are $I_{s1}|\sin\theta|$ and $I_{s2}$ calculated from Eqs. \eqref{eq2_I1} and \eqref{eq2_I2}, respectively. The factor of $|\sin\theta|$ arising from the rotation of YIG1 with respect to the magnetic field {\bf H} due to the ISHE in the Pt layer.
The agreement between experiment and model indicates that our long distance manipulation of spin current in YIG2 is due to the cooperativity control of YIG1. 
Therefore control of the cooperativities allows us to control the dispersion of 
the strongly coupled system and thereby directly and remotely control the spin current amplitude in both samples simultaneously.

In summary, we have electrically detected spin currents from two YIG/Pt samples which both couple to a cavity mode.
Via such a local detection technique we are able to distinguish the spin dynamics in each sample individually and demonstrate the manipulation of distant spin currents, whereby controlling the cooperativity of one magnetic sample will manipulate the spin current in another sample well separated from the first.  Such a long distance manipulation originates in the local spin current dependence on the global coupling properties and the ability to locally detect the spin system through electrical detection.  By demonstrating and explaining such long distance manipulation this work opens a new avenue to spin current generation and manipulation techniques in the developing field of cavity spintronics.  

M.H. is partially supported by an NSERC CGSD Scholarship and IODE Canada. 
P.H. is supported by the UMGF program.
This work has been funded by NSERC, CFI, NSFC (11429401) grants (C.-M.H.) and NSF DMR 1505192 (J.Q.X.).

  \bibliography{bai.bbl}

\end{document}